\documentclass[12pt,reqno]{article}
\usepackage{amsfonts}
\usepackage{amsmath}
\usepackage{amsbsy}

\usepackage{amssymb,latexsym}

 \numberwithin{equation}{subsection}

\begin{document}
 \allowdisplaybreaks[1]
\title{Supergravity Induced Interactions on Thick Branes}
\author{Nejat T. Y$\i$lmaz\\
Department of Electrical and Electronics Engineering,\\
Ya\c{s}ar University,\\
Sel\c{c}uk Ya\c{s}ar Kamp\"{u}s\"{u}\\
\"{U}niversite Caddesi, No:35-37,\\
A\u{g}a\c{c}l\i Yol, 35100,\\
Bornova, \.{I}zmir, Turkey.\\
\texttt{nejat.yilmaz@yasar.edu.tr}} \maketitle
\begin{abstract}
The gravity coupling of the symmetric space sigma model is studied
in the solvable Lie algebra parametrization. The corresponding
Einstein's equations are derived and the energy-momentum tensor is
calculated. The results are used to derive the dynamical equations
of the warped $5D$ geometry for localized bulk scalar interactions
in the framework of thick brane world models. The Einstein and
scalar field equations are derived for flat brane geometry in the
context of minimal and non-minimal gravity-bulk scalar couplings.
\end{abstract}

\section{Introduction}
Brane world scenarios which have their roots in open string theory
ingredients $D$-branes are alternative approaches to relate higher
dimensions to the standard  model of fundamental interactions.
Instead of considering diminished and compact extra dimensions
they provide sensible resolution especially to the hierarchy
problem with non-compact extra dimensions. Although there were
pioneer ideas in the 1980's \cite{1,2} the string theory inspired
brane world cosmological scenarios were proposed and constructed
much later \cite{31,32,33,34,35,36,37,38}. These theories have
flat or non-flat (warped), compact or non-compact bulk geometries.
However they share the common feature that the $4D$ spacetime is
an embedded solution in the bulk in such a way that the standard
model fields are localized on it where as gravity can probe the
extra dimension(s). The foundation of this picture lies in the
dynamics of the various superstring theories in which the charge
carrier open strings may be described by their end point dynamics
of D-branes whereas the closed strings which accommodate gravitons
are not constrained to live on the brane like open string
boundaries. Unlike the pioneer models which ignore the brane size
by explicitly containing Dirac-delta functions more realistic
brane models which have finite brane size have been also
constructed and are currently being studied
\cite{41,42,43,44,45,46,47,48,49,410,411,412,413,414,415,416,417,temel}.
These models fulfill the requirement of a fundamental length scale
due to string theory. An updated review of these so-called thick
branes can be found in \cite{thick}

In \cite{temel} it is discussed that the localized interactions of
bulk scalars on a thick brane solution of a warped $5D$ bulk can
be described by a sigma-model in which the brane forming scalar is
non-linearly coupled to the bulk scalar fields via its
derivatives. In this manner the kinetic term of the brane forming
scalar is a function of the bulk scalars.

On the other hand the non-linear sigma model \cite{sig1,sig2,sig3}
governs the scalar sector of the supergravity theories
\cite{westbook,sssugradivdim,westsugra,tani}. In particular if the
target space (the scalar manifold) is a symmetric space \cite{hel}
then we have a special case of the general sigma model which can
be named as the symmetric space sigma model \cite{sm1,sm2,sm3}.
The compactifications of the $D=11$ supergravity (with its
$S^{1}$-compactified IIA supergravity redundant) as well as the
$D=10$ IIB supergravity and the $D=10$ type I supergravity which
is coupled to the Yang-Mills theory in majority produce scalar
sectors in lower dimensions in the form of symmetric space sigma
models. Moreover the IIA and the IIB supergravities describe the
tree-level low energy dynamics of the type IIA and the type IIB
superstring theories whereas the type I Yang-Mills supergravity is
the effective low energy limit of the heterotic superstring
theory. With this motivation in the above mentioned context of
\cite{temel} in this work we study and derive the thick brane
dynamical equations in the presence of a generic symmetric space
sigma model coupling of the $5D$ bulk gravity. We will focus on
the flat Minkowski brane dynamics and derive the corresponding
Einstein and scalar field equations for the warped geometry of the
bulk. We will consider the both cases of the minimal and the
non-minimal gravity-scalar coupling. Our formulation will be for
localized bulk scalars coupling to the brane solution-generating
scalar via a generic symmetric space sigma model lagrangian
constructed in the solvable lie algebra gauge. The outline of the
construction of symmetric space sigma models by means of the
solvable Lie algebra or the axion-dilaton parametrization can be
referred in \cite{julia1,julia2,nej1,nej2,nej3,symmspace}.
Furthermore in \cite{sssm1} an explicit construction of the
symmetric space sigma model lagrangian in terms of the solvable
Lie algebra parameters of the target space is presented for
arbitrary trace conventions. The most general form of the field
equations are also derived for the target space coordinates in the
same work within a general algebraic formalism.

In Section two we consider the coupling of the gravity to the
symmetric space sigma model in a dimension-free and general
framework. After defining the action we derive the field equations
and identify the energy-momentum tensor which we will adopt in the
following section. In Section three we turn our attention to the
dynamics which give rise to thick branes with localized bulk
scalar interactions on them. We will first discuss the ingredients
of the warped $5D$ geometry then we will derive the corresponding
dynamical equations for the minimal and non-minimal gravity-scalar
couplings. We will also express these equations in their
appropriate form for a first-order formalism. Finally we will
obtain the scalar field equations in the $5D$ warped geometry
context. For completeness we collect some of the variational steps
followed in deriving field equations of gravity-scalar couplings
in the Appendix.
\section{Coupling gravity to the symmetric space $\sigma$-model}\label{section1}
In this section we will focus on the gravity coupling of the
scalar sectors of the dimensionally-reduced supergravity theories
which can in general be modelled as symmetric space or reductive
coset sigma models \cite{sm1,sm2,sm3}. Our construction will be
for a general $G/K$ coset scalar manifold where the scalars live
in $D$-dimensions. In the next section we will use our general
results in the context of thick brane world scenarios in which the
scalar configuration will correspond to the interactions on the
brane.
\subsection{\textsl{The action}}
 The general non-linear sigma model \cite{sig1,sig2,sig3}
action can be given as
\begin{equation}\label{e1}
 S_{NLSM}=\int d^{(D)}\sigma\sqrt{-h}\:h^{AB}
 g_{ab}\partial_{A}\varphi^{a}\partial_{B}\varphi^{b}.
\end{equation}
To clarify \eqref{e1} one has to consider an immersion map
\begin{equation}\label{e2}
f:\:N\longrightarrow M,
\end{equation}
of a smooth $D$-dimensional manifold $N$ into another smooth
manifold $M$. In supergravity $N$ is the spacetime and $M$ is the
scalar manifold on the other hand in $p$-brane dynamics $N$ is the
world volume of the $p$-brane and $M$ is the background space. The
action \eqref{e1} is defined on $N$. If one considers the metric
$h_{AB}$ on $N$ as an independent field and varies the action
\eqref{e1} with respect to it one finds the corresponding field
equations
\begin{equation}\label{e3}
 h_{AB}=g_{ab}\partial_{A}\varphi^{a}\partial_{B}\varphi^{b},
\end{equation}
which denote that $h_{AB}$ comes out to be the pullback of the
metric $g_{ab}$ which is defined on $M$ onto $N$ through the
immersion map \eqref{e2} of $N$. In \eqref{e1} $h$ is the
determinant of the metric $h_{AB}$. If on a chart of $M$ the
coordinates of the target space $M$ are taken to be $\varphi^{b}$
then they can be considered to be scalar fields on $N$ via the
immersion map \eqref{e2}. For a local coordinate chart of $N$ with
coordinates $x^{A}$ composition of the coordinate charts with
\eqref{e2} also gives the fields $\varphi^{b}(x^{A})$ which appear
in \eqref{e1}. In particular the target manifold $M$ can be chosen
to be a coset space $G/K$. Furthermore if $G$ is a non-compact
real form of any other semi-simple Lie group and if $K$ is a
maximal compact subgroup of $G$, also if the Lie algebra of $K$ is
a maximal compactly imbedded Lie subalgebra of the Lie algebra of
$G$ then $G/K$ is a Riemannian globally symmetric space for all
the $G$-invariant Riemannian structures on it \cite{hel}. In this
case the sigma model is called the symmetric space sigma model
(SSSM) \cite{sm1,sm2,sm3}. The action of the SSSM can be given as
\cite{julia1,julia2,nej1,nej2,nej3,symmspace,sssm1}
\begin{equation}\label{e4}
 S_{SSSM}=\frac{1}{4}\int tr(\ast d{\mathcal{M}}^{-1}\wedge
 d{\mathcal{M}}),
\end{equation}
where $\mathcal{M}$ is a map from $N$ into $M=G/K$ which is based
on the global solvable Lie algebra parametrization of $G/K$ and
the trace is over the matrix representation of $G$.\footnote{We
refer the reader to the references given above for a detailed
construction of \eqref{e4}.} \eqref{e4} is invariant under the
global action of $G$ from the right and the local action of $K$
from the left. In \cite{sssm1} \eqref{e4} is explicitly derived in
terms of the global solvable Lie algebra parameters of $G/K$
\cite{julia1,julia2,nej1,nej2,nej3,symmspace,sssm1} namely in
terms of the fields
\begin{equation}\label{e5}
  \{\phi^{1},
  \phi^{2},
  \cdots,
  \phi^{r},
  \chi^{1},
  \chi^{2},
  \cdots,
  \chi^{n}\},
\end{equation}
which can be considered to be independent scalar fields on $N$ and
for which $ r+n=\text{dim}(G/K)$.\footnote{In the following we
will use the indices $i,j,...$ for the fields $\{\phi^{i}\}$, the
indices $\alpha,\beta,\gamma,...$ for the fields
$\{\chi^{\beta}\}$, and we will use $A,B,C,...$ as the indices of
the manifold $N$.} We may further assume that the global
parametrization of \eqref{e5} coincides with a local coordinate
chart of $M$. Also if we take a local chart for $N$ then the
fields in $\eqref{e5}$ can be considered to be coinciding with the
ones in \eqref{e1}. In terms of these scalar fields from
\cite{sssm1} the action reads
\begin{subequations}\label{e6}
\begin{align}
S_{SSSM}=\int\bigg(&-\frac{1}{8}\,\mathcal{A}_{ij}\ast
d{\phi}^{i}\wedge
 d\phi^{j}-\frac{1}{4}\,\mathcal{B}_{i\alpha}\ast d{\phi}^{i}\wedge
e^{\frac{1}{2}\alpha_{j}\phi^{j}} \mathbf{\Omega}^{\alpha}_{\gamma}d\chi^{\gamma}\notag\\
&-\frac{1}{2}\,\mathcal{C}_{\alpha\beta}e^{\frac{1}{2}\alpha_{j}\phi^{j}}\ast
\mathbf{\Omega}^{\alpha}_{\gamma}d\chi^{\gamma}\wedge
e^{\frac{1}{2}\beta_{i}\phi^{i}}
\mathbf{\Omega}^{\beta}_{\tau}d\chi^{\tau}\bigg) ,\tag{\ref{e6}}
\end{align}
\end{subequations}
where the coefficients
$\mathcal{A}_{ij},\mathcal{B}_{i\alpha},\mathcal{C}_{\alpha\beta}$
are normalization constants which originate from the choice of the
matrix representation of $G$, their exact definitions may be found
in \cite{sssm1}. $\alpha_{j}$ and $\beta_{i}$ are the root vector
components of the Cartan generators coupled to the fields
$\phi^{i}$ in the solvable Lie algebra parametrization of $G/K$.
We will not explicitly construct the solvable Lie algebra
parametrization of the symmetric space $G/K$ here which is
extensive in its own right therefore we again refer the reader to
the references \cite{nej1,nej2,nej3,symmspace} for further
definitions where the solvable Lie algebra gauge is studied in
detail. For our purposes we will also not need to know the
explicit form of the $n\times n$ matrix functions
\begin{equation}\label{e7}
\mathbf{\Omega}= \mathbf{\Omega}(\chi^{\beta}),
\end{equation}
which appear in \eqref{e6} and which will solely appear in our
further formulation as coefficient functions. For this reason we
also refer the reader to the references
\cite{nej1,nej2,nej3,symmspace} for their rigorous derivation and
the involved definitions. Next we will introduce the elements of
the gravity sector which we will couple to \eqref{e4}. Since brane
world scenarios are open to non-standard and modified gravity
formalisms we will try to be specific and transparent in our
definitions. The gravity coupling will be by means of the metric
$h_{AB}$ on $N$. Thus we will consider the unique
metric-compatible, torsion-free Levi-Civita connection of the
pseudo-Riemannian structure $h_{AB}$ on $N$. The gravitational
sector on $N$ reads
\begin{equation}\label{e8}
 S_{GRAV}=\int R_{AB}\wedge\ast e^{AB},
\end{equation}
where we have chosen an orthogonal moving co-frame $\{e^{A}\}$
with $A=1,...,D$ on $N$. The curvature two-forms $R_{AB}$ of the
unique Levi-Civita connection of $h_{AB}$ can be written as
\footnote{We should state that we raise and lower indices by using
the metric $h_{AB}$. Since we choose an orthogonal moving co-frame
the metric components $\{h_{AB}\}$ are constant thus one can
freely raise and lower indices on both sides of \eqref{e9}. Also
in particular we assume that $N$ is connected which guarantees
that the signature of the pseudo-Riemannian structure $h_{AB}$ is
constant on $N$.}
\begin{equation}\label{e9}
R_{AB}=d\omega_{AB}+\omega_{AC}\wedge \omega^{C}_{\:\:\:B},
\end{equation}
where $\{\omega^{A}_{\:\:\: B}\}$ are the connection one-forms of
the Levi-Civita connection of $h_{AB}$ generated by the moving
co-frame $\{e^{A}\}$. Since the moving co-frame is an orthogonal
one the metric-compatibility reads
\begin{equation}\label{e10}
\omega_{AB}=-\omega_{BA}.
\end{equation}
The torsion-free condition can be written as
\cite{r1,r2,r3,thring,tucker,nakahara}
\begin{equation}\label{e11}
de^{A}=-\omega^{A}_{\:\:\: B}\wedge e^{B}.
\end{equation}
We may also introduce a potential term for the scalar fields as
\begin{equation}\label{e11.5}
 S_{POT}=\int \ast V,
\end{equation}
where $V=V(\phi^{i},\chi^{\beta})$ is the potential which usually
arises as a result of gauging away some of the symmetry in the
scalar sector. Now we can write the total action as
\begin{equation}\label{e11.6}
 S=\int\bigg(R_{AB}\wedge\ast e^{AB}-\ast
V-\frac{1}{4}tr(\ast d{\mathcal{M}}^{-1}\wedge
 d{\mathcal{M}})\bigg),
\end{equation}
where we have chosen the negative sign in front of the sigma model
term to achieve positive kinetic terms for the scalars. Explicitly
we have
\begin{subequations}\label{e12}
\begin{align}
S=\int\bigg(&R_{AB}\wedge\ast e^{AB}-\ast
V+\frac{1}{8}\,\mathcal{A}_{ij}\ast d{\phi}^{i}\wedge
 d\phi^{j}
 +\frac{1}{4}\,\mathcal{B}_{i\alpha}\ast d{\phi}^{i}\wedge
e^{\frac{1}{2}\alpha_{j}\phi^{j}}
\mathbf{\Omega}^{\alpha}_{\gamma}d\chi^{\gamma}\notag\\
&+\frac{1}{2}\,\mathcal{C}_{\alpha\beta}e^{\frac{1}{2}\alpha_{j}\phi^{j}}\ast
\mathbf{\Omega}^{\alpha}_{\gamma}d\chi^{\gamma}\wedge
e^{\frac{1}{2}\beta_{i}\phi^{i}}
\mathbf{\Omega}^{\beta}_{\tau}d\chi^{\tau}\bigg). \tag{\ref{e12}}
\end{align}
\end{subequations}
The variation of this action results in the same field equations
for the scalar fields with the ones already derived in
\cite{sssm1} for the pure symmetric space sigma model. Essentially
the metric $g$ on $M$ is responsible for the global and the local
symmetry of the sigma model action \eqref{e4}. There is a
remarkable difference between the pure and the gravity coupled
case. In the pure sigma model the field equation of the the metric
$h$ is simply \eqref{e3} which denotes that the metric $h$ on $N$
is the one induced by $g$ via \eqref{e2}. Thus since $h$ is
considered to be an independent field, to generate the solution
space of the model one may choose an arbitrary $h$ and then solve
the sigma model field equations which will minimize the pure sigma
model action and which will also indirectly determine the metric
$g$. One may repeat this procedure by changing $h$ to generate the
entire solution space. Coupling the sigma model to the gravity on
the other hand abolishes this methodology. Now the coupling occurs
by means of $h$ and $h$ can not be chosen arbitrarily since when
we introduce gravity on $N$ we solve for the class of
pseudo-Riemannian metrics $h$ whose curvatures minimize the action
\eqref{e12}. The field equations of the scalar fields contain $h$
whereas the Einstein's equations as it will be clear in the next
subsection contain the scalar fields as sources. Thus these two
sets of equations must be solved simultaneously. Another
divergence between the two cases is that the energy-momentum
tensor associated with the fields \eqref{e5} and the potential
will not be null in the gravity coupled case thus \eqref{e3} is no
longer valid and the metric $h$ on $N$ is not induced by the
metric $g$ on $M$ via \eqref{e2} in this case.
\subsection{\textsl{The field equations}}
 In this subsection we will vary the action \eqref{e12} to obtain the
corresponding field equations of the SSSM that is coupled to
gravity with also a potential term. We will not lay out the
detailed steps of this variation, and we have collected some of
the results in the Appendix. We refer the reader to \cite{thring}
for the standard elements of the variational calculus of
differential forms. We will divide the variation of the action
into two parts; the first containing the gravitational variation
terms whereas the second containing the variation of the scalars
introduced in \eqref{e5} which is independent of the variation of
the metric. Thus we have
\begin{equation}\label{e13}
\delta S=\delta S_{1}+\delta S_{2}.
\end{equation}
For the Levi-Civita connection and for the choice of an orthogonal
moving co-frame $\{e^{A}\}$ the explicit form of the first term
which is induced by the variation of the orthogonal co-frame can
be seen in the Appendix. By imposing the least action principle in
\eqref{e13} explicitly from \eqref{app1} we can obtain the
Einstein's equations which read
\begin{subequations}\label{e18}
\begin{align}
\ast e_{ABC}\wedge R^{AB}=&\:\:\:V\ast e_{C}
+(-1)^{D-1}\frac{1}{8}\,\mathcal{A}_{ij}(d{\phi}^{i}\wedge i_{C}
\ast d\phi^{j} +i_{C}d{\phi}^{j}\wedge\ast
d\phi^{i})\notag\\
&+(-1)^{D-1}\frac{1}{4}\,\mathcal{B}_{i\alpha}
e^{\frac{1}{2}\alpha_{j}\phi^{j}}
\mathbf{\Omega}^{\alpha}_{\beta}(d{\phi}^{i}\wedge i_{C} \ast
d\chi^{\beta} +i_{C}d\chi^{\beta}\wedge\ast d\phi^{i})\notag\\
&+(-1)^{D-1}\frac{1}{2}\,\mathcal{C}_{\alpha\beta}
e^{\frac{1}{2}\alpha_{i}\phi^{i}}e^{\frac{1}{2}\beta_{j}\phi^{j}}
\mathbf{\Omega}^{\alpha}_{\tau}\mathbf{\Omega}^{\beta}_{\rho}(d\chi^{\tau}\wedge
i_{C} \ast d\chi^{\rho}\notag\\
 &+i_{C}d\chi^{\rho}\wedge\ast
d\chi^{\tau}), \tag{\ref{e18}}
\end{align}
\end{subequations}
where we have introduced the interior derivative
$i_{\omega_{1}}\omega_{2}$ \cite{thring}
\begin{equation}\label{e15}
i\: :\: (\omega_{1},\omega_{2})\in E_{p}\times
E_{q}\longrightarrow i_{\omega_{1}}\omega_{2}\in E_{(q-p)},
\end{equation}
which takes a $p$-form and a $q$-form and maps them to a
$(q-p)$-form. We use the notation $i_{C}\equiv i_{e^{C}}$. The
second term in \eqref{e13} which contains the non-gravitational
variation of the fields is also derived in the Appendix. Equating
it separately to zero leads us to the scalar field equations. From
\eqref{app3} we read the dilatonic scalar field equations as
\begin{equation}\label{e15.5}
\begin{aligned}
(-1)^{(D-1)}&d(\frac{1}{2}(\mathcal{A}_{ik}+\mathcal{A}_{ki})\ast
d\phi ^{i}+\mathcal{B}_{k\alpha}e^{\frac{1}{2}\alpha _{i}\phi
^{i}}\mathbf{\Omega}^{\alpha}_{\beta}\ast
d\chi^{\beta})\\
\\
=&\frac{1}{2}\mathcal{B}_{i\alpha}\alpha_{k}\ast d\phi ^{i}\wedge
e^{\frac{1}{2}\alpha _{i}\phi
^{i}}\mathbf{\Omega}^{\alpha}_{\beta}d\chi^{\beta}\\
\\
&+\mathcal{C}_{\alpha\beta}(\alpha_{k}+\beta_{k})e^{\frac{1}{2}\alpha
_{i}\phi ^{i}}\mathbf{\Omega}^{\alpha}_{\tau}\ast
d\chi^{\tau}\wedge e^{\frac{1}{2}\beta _{j}\phi
^{j}}\mathbf{\Omega}^{\beta}_{\gamma}d\chi^{\gamma}-4\partial_{k}V\ast
1.
\end{aligned}
\end{equation}
Similarly again via \eqref{app3} we obtain the axionic scalar
field equations as
\begin{equation}\label{e15.6}
\begin{aligned}
(-1)^{(D-1)}&d(\frac{1}{2}\mathcal{B}_{i\alpha}e^{\frac{1}{2}\alpha
_{i}\phi ^{i}}\mathbf{\Omega}^{\alpha}_{\theta}\ast d\phi
^{i}+\mathcal{C}_{\alpha\beta}e^{\frac{1}{2}\alpha _{i}\phi
^{i}}e^{\frac{1}{2}\beta_{j}\phi
^{j}}(\mathbf{\Omega}^{\alpha}_{\gamma}\mathbf{\Omega}^{\beta}_{\theta}+
\mathbf{\Omega}^{\alpha}_{\theta}\mathbf{\Omega}^{\beta}_{\gamma})\ast
d\chi^{\gamma})\\
\\
=&\frac{1}{2}\mathcal{B}_{i\alpha}\mathcal{D}^{\alpha}_{\theta\beta}e^{\frac{1}{2}\alpha
_{i}\phi ^{i}}\ast d\phi ^{i}\wedge d\chi^{\beta}\\
\\
&+\mathcal{C}_{\alpha\beta}e^{\frac{1}{2}\alpha _{i}\phi
^{i}}e^{\frac{1}{2}\beta _{j}\phi
^{j}}(\mathcal{D}^{\alpha}_{\theta\tau}\mathbf{\Omega}^{\beta}_{\gamma}+
\mathbf{\Omega}^{\alpha}_{\tau}\mathcal{D}^{\beta}_{\theta\gamma})\ast
d\chi^{\tau}\wedge d\chi^{\gamma}-2\partial_{\theta}V\ast 1,
\end{aligned}
\end{equation}
where as we have introduced in the Appendix the matrix functions
$\mathcal{D}_{\theta}\equiv\frac{\partial\mathbf{\Omega}}{\partial\chi^{\theta}}$.
in \eqref{e15.5} and \eqref{e15.6} are already derived in
\cite{sssm1} by a direct application of the Euler-Lagrange
equations to the pure scalar action. However in the present work
in the Appendix we have preferred to obtain a complete formulation
for the variation of the total action.
\subsection{\textsl{The energy-momentum tensor}}
Now following the identification of the energy-momentum one-forms
associated with the symmetric space sigma model and the
corresponding scalar potential coupling of the pure gravity we
will derive the component expression of the energy-momentum tensor
resulting from these sources in \eqref{e12}. Starting with the
Einstein's equations \eqref{e18} we immediately see that the
energy-momentum one-forms $t_{C}$ satisfy \cite{thring}
\begin{subequations}\label{e21}
\begin{align}
\ast t_{C}=&- V\ast e_{C}
-(-1)^{D-1}\frac{1}{8}\,\mathcal{A}_{ij}(d{\phi}^{i}\wedge i_{C}
\ast d\phi^{j} +i_{C}d{\phi}^{j}\wedge\ast
d\phi^{i})\notag\\
&-(-1)^{D-1}\frac{1}{4}\,\mathcal{B}_{i\alpha}
e^{\frac{1}{2}\alpha_{j}\phi^{j}}
\mathbf{\Omega}^{\alpha}_{\beta}(d{\phi}^{i}\wedge i_{C} \ast
d\chi^{\beta} +i_{C}d\chi^{\beta}\wedge\ast d\phi^{i})\notag\\
&-(-1)^{D-1}\frac{1}{2}\,\mathcal{C}_{\alpha\beta}
e^{\frac{1}{2}\alpha_{i}\phi^{i}}e^{\frac{1}{2}\beta_{j}\phi^{j}}
\mathbf{\Omega}^{\alpha}_{\tau}\mathbf{\Omega}^{\beta}_{\rho}(d\chi^{\tau}\wedge
i_{C} \ast d\chi^{\rho}\notag\\
 &+i_{C}d\chi^{\rho}\wedge\ast
d\chi^{\tau}). \tag{\ref{e21}}
\end{align}
\end{subequations}
Now we may express the field strengths of the scalar fields
$\{\phi^{i},\chi^{\beta}\}$ in terms of their components with
respect to the orthogonal moving co-frame $\{e^{C}\}$ on $N$ as
\begin{subequations}\label{e22}
\begin{align}
 d\phi^{i}&=\mathcal{F}^{i}_{\:\:C}e^{C},\notag\\
 \tag{\ref{e22}}\\
 d\chi^{\beta}&=\mathcal{H}^{\beta}_{\:\:C}e^{C}.\notag
\end{align}
\end{subequations}
If the orthogonal moving co-frame $\{e^{C}\}$ is taken to coincide
with a coordinate basis $\{dx^{C}\}$ then we have
\begin{subequations}\label{e22.5}
\begin{align}
 d\phi^{i}&=\frac{\partial\phi^{i}(x^{A})}{\partial x^{C}}dx^{C}\equiv\partial_{C}\phi^{i}dx^{C},\notag\\
 \tag{\ref{e22.5}}\\
 d\chi^{\beta}&=\frac{\partial\chi^{\beta}(x^{A})}{\partial x^{C}}dx^{C}\equiv\partial_{C}\chi^{\beta}dx^{C},\notag
\end{align}
\end{subequations}
which implies that
\begin{subequations}\label{e22.6}
\begin{align}
 \mathcal{F}^{i}_{\:\:C}&=\frac{\partial\phi^{i}(x^{A})}{\partial x^{C}}\equiv\partial_{C}\phi^{i}(x^{A}),\notag\\
 \tag{\ref{e22.6}}\\
 \mathcal{H}^{\beta}_{\:\:C}&=\frac{\partial\chi^{\beta}(x^{A})}{\partial x^{C}}\equiv\partial_{C}\chi^{\beta}(x^{A}).\notag
\end{align}
\end{subequations}
By taking the Hodge-dual of both sides of \eqref{e21}, also by
inserting the component expansions \eqref{e22} and by further
simplifying we can obtain the energy-momentum one-forms as
\begin{subequations}\label{e23}
\begin{align}
t_{C}=\bigg(&-V h_{CA}
-(-1)^{D-1}\bigg(\frac{1}{8}\,\mathcal{A}_{ij}(-\mathcal{F}^{iB}\mathcal{F}^{j}_{\:\:B}h_{CA}
+\mathcal{F}^{i}_{\:\:C}\mathcal{F}^{j}_{\:\:A}+\mathcal{F}^{j}_{\:\:C}\mathcal{F}^{i}_{\:\:A})\notag\\
&+\frac{1}{4}\,\mathcal{B}_{i\alpha}
e^{\frac{1}{2}\alpha_{j}\phi^{j}}
\mathbf{\Omega}^{\alpha}_{\beta}(-\mathcal{F}^{iB}\mathcal{H}^{\beta}_{\:\:B}h_{CA}
+\mathcal{F}^{i}_{\:\:C}\mathcal{H}^{\beta}_{\:\:A}+\mathcal{H}^{\beta}_{\:\:C}\mathcal{F}^{i}_{\:\:A})\notag\\
&+\frac{1}{2}\,\mathcal{C}_{\alpha\beta}
e^{\frac{1}{2}\alpha_{i}\phi^{i}}e^{\frac{1}{2}\beta_{j}\phi^{j}}
\mathbf{\Omega}^{\alpha}_{\tau}\mathbf{\Omega}^{\beta}_{\rho}
(-\mathcal{H}^{\tau B}\mathcal{H}^{\rho}_{\:\:B}h_{CA}
+\mathcal{H}^{\tau}_{\:\:C}\mathcal{H}^{\rho}_{\:\:A}\notag\\
&+\mathcal{H}^{\rho}_{\:\:C}\mathcal{H}^{\tau}_{\:\:A})\bigg)\bigg)e^{A}.
\tag{\ref{e23}}
\end{align}
\end{subequations}
Since the energy-momentum tensor components $\{T_{CA}\}$ are
defined through
\begin{equation}\label{e24}
t_{C}=T_{CA}e^{A},
\end{equation}
we can finally write the energy-momentum tensor in terms of the
components of the scalar field strengths which are expanded in the
chosen orthogonal moving co-frame as

\begin{equation}\label{e24.5}
T_{CA}=\:\:\:-Vh_{CA}+\mathcal{T}_{CA},
\end{equation}
where
\begin{subequations}\label{e25}
\begin{align}
\mathcal{T}_{CA}=
&-(-1)^{D-1}\bigg(\frac{1}{8}\,\mathcal{A}_{ij}(-\mathcal{F}^{iB}\mathcal{F}^{j}_{\:\:B}h_{CA}
+\mathcal{F}^{i}_{\:\:C}\mathcal{F}^{j}_{\:\:A}+\mathcal{F}^{j}_{\:\:C}\mathcal{F}^{i}_{\:\:A})\notag\\
&+\frac{1}{4}\,\mathcal{B}_{i\alpha}
e^{\frac{1}{2}\alpha_{j}\phi^{j}}
\mathbf{\Omega}^{\alpha}_{\beta}(-\mathcal{F}^{iB}\mathcal{H}^{\beta}_{\:\:B}h_{CA}
+\mathcal{F}^{i}_{\:\:C}\mathcal{H}^{\beta}_{\:\:A}+\mathcal{H}^{\beta}_{\:\:C}\mathcal{F}^{i}_{\:\:A})\notag\\
&+\frac{1}{2}\,\mathcal{C}_{\alpha\beta}
e^{\frac{1}{2}\alpha_{i}\phi^{i}}e^{\frac{1}{2}\beta_{j}\phi^{j}}
\mathbf{\Omega}^{\alpha}_{\tau}\mathbf{\Omega}^{\beta}_{\rho}
(-\mathcal{H}^{\tau B}\mathcal{H}^{\rho}_{\:\:B}h_{CA}
+\mathcal{H}^{\tau}_{\:\:C}\mathcal{H}^{\rho}_{\:\:A}+\mathcal{H}^{\rho}_{\:\:C}\mathcal{H}^{\tau}_{\:\:A})\bigg),
\tag{\ref{e25}}
\end{align}
\end{subequations}
is the energy-momentum tensor contribution of the symmetric space
sigma model. In the next Section when we consider the thick brane
world scenario we will adopt especially the sigma model part of
the energy-momentum tensor \eqref{e24.5}. However after renaming
the coefficients it will be reduced to a simpler form by choosing
the co-frame  as a coordinate one and in particular by assigning a
metric ansatz for the manifold $N$.
\section{Bulk scalar interactions on thick $\\$ brane worlds}\label{section2}
After a detailed analysis of the symmetric space sigma model-
gravity coupling in the previous section we will now turn our
attention to the main objective of the present work which aims the
derivation of the dynamical equations describing the bulk scalar
interactions within the context of the smoothed brane world
scenario. As discussed in \cite{temel} the bulk scalars interact
with the brane-forming scalar within the framework of a
sigma-model in which the bulk moduli scalars couple to the kinetic
term of the brane-forming one. As we have already discussed in the
Introduction we will focus on the symmetric space sigma model form
of interactions which is the most realistic one as it appears as a
result of the dimensional reduction of the higher dimensional
supergravity theories.

The link between the physical scenarios discussed in this section
and the formal gravity-symmetric space sigma model coupling of the
previous section will be apparent when we consider the minimal
coupling in sub-section 3.2. In summary in the gravity-sigma model
coupling context we will assume a $5D$ warped geometry of the bulk
which accommodates a Minkovski brane with a warp factor and derive
the Einstein and scalar field equations for the $5D$ bulk in
component form (rather than their global form of the previous
section) so that they will lead us to the modified $4D$ brane
spacetime dynamics which is implicit in the warped-geometric
cosmological scenario. The formalism of minimal coupling in
sub-section 3.2 is almost the same with Section two and in the
non-minimal coupling of sub-section 3.3 we will modify this
formalism by deforming the Einstein-Hilbert term. The reason why
we have separately studied the sigma model-gravity coupling in
Section two is basically to derive the right hand side of the
Einstein equations namely the matter energy-momentum tensor
arising from the scalars of the sigma model in a neat and a
general framework. On the other hand in this section we will work
out the left hand side of the Einstein equations namely the
Einstein tensor for the special geometry mentioned above. For both
of the minimal and the non-minimal couplings we will adopt the
energy-momentum tensor of the sigma model generically derived in
Section two.
\subsection{\textsl{5D geometry}}
Before we discuss the actions which define the brane dynamics in
the presence of bulk scalars let us set the back-scene geometry.
First of all we fix the bulk dimension to five so that $dimN=5$
for the notation we have introduced in the previous section. In
this work we will consider the flat (Minkovski) geometry for the
brane \cite{temel,thick,tem1,tem2}. Therefore our $5D$ metric
ansatz will be
\begin{equation}\label{e30}
h=e^{2A}\eta_{\mu\nu}dx^{\mu}\otimes dx^{\nu}-dy\otimes dy,
\end{equation}
where $x^{4}=y$ is the fifth dimension coordinate, $A=A(y)$, and
$e^{2A}$ is the warp-factor for the brane geometry. Our
conventions are such that: $A,B,C=0,1,2,3,4$; $\mu,\nu=0,1,2,3$;
and $\eta_{\mu\nu}=diag(1,-1,-1,-1)$. Now considering the $5D$
metric $h_{AB}$ and its inverse $h^{AB}$ since via \eqref{e30} we
have
\begin{subequations}\label{e31}
\begin{align}
h&=\text{diag}(e^{2A},-e^{2A},-e^{2A},-e^{2A},-1),\notag\\
h^{-1}&=\text{diag}(e^{-2A},-e^{-2A},-e^{-2A},-e^{-2A},-1),
\tag{\ref{e31}}
\end{align}
\end{subequations}
from
\begin{equation}\label{e32}
\Gamma^{A}_{\:\:\:BC}=\frac{1}{2}h^{DA}(h_{DB,C}+h_{DC,B}-h_{BC,D}),
\end{equation}
one can calculate the Levi-Civita connection coefficients for the
coordinate frame suggested in \eqref{e30}. They read
\begin{equation}\label{e33}
\begin{aligned}
\Gamma^{4}_{\:\:\:44}=\Gamma^{4}_{\:\:\:A4}=\Gamma^{A}_{\:\:\:44}&=\Gamma^{\mu}_{\:\:\:44}=\Gamma^{\gamma}_{\:\:\:\mu\nu}=0,\\
\\
\Gamma^{\mu}_{\:\:\:\nu4}=A^{\prime}\delta^{\mu}_{\:\:\:\nu},\quad
&\Gamma^{4}_{\:\:\:\mu\nu}=e^{2A}A^{\prime}\eta_{\mu\nu},
\end{aligned}
\end{equation}
where prime denotes derivation with respect to the coordinate $y$.
Via the Riemann tensor
\begin{equation}\label{e34}
R^{A}_{\:\:\:BCD}=\Gamma^{A}_{\:\:\:BD,C}-\Gamma^{A}_{\:\:\:BC,D}+\Gamma^{A}_{\:\:\:EC}\Gamma^{E}_{\:\:\:BD}-\Gamma^{A}_{\:\:\:ED}\Gamma^{E}_{\:\:\:BC},
\end{equation}
we can calculate the Ricci tensor $R_{AB}=R^{C}_{\:\:\:ACB}$
components and the Ricci scalar $R=h^{AB}R_{AB}$ as
\begin{equation}\label{e35}
\begin{aligned}
R_{4\mu}=0,\quad R_{44}=-4&(A^{\prime})^{2}-4A^{\prime\prime},\\
\\
R_{\mu\nu}=e^{2A}(4(A^{\prime})^{2}+A^{\prime\prime})\eta_{\mu\nu},&\quad
R=20(A^{\prime})^{2}+8A^{\prime\prime}.
\end{aligned}
\end{equation}
Now we can calculate the corresponding Einstein tensor components
$G_{AB}=R_{AB}-\frac{1}{2}h_{AB}R$ as
\begin{equation}\label{e36}
\begin{aligned}
G_{4\mu}=0,\quad &G_{44}=6(A^{\prime})^{2},\\
\\
G_{\mu\nu}=e^{2A}(-6&(A^{\prime})^{2}-3A^{\prime\prime})\eta_{\mu\nu}.
\end{aligned}
\end{equation}
Furthermore for a generic scalar field $f=f(y)$ by considering the
double action of the covariant derivative
\begin{equation}\label{e36.5}
\nabla_{A}\nabla_{B}f=\partial_{A}\partial_{B}f-\partial_{C}f\Gamma^{C}_{\:\:\:BA},
\end{equation}
via \eqref{e33} one can show that
\begin{equation}\label{e36.6}
\begin{aligned}
\nabla_{\mu}\nabla_{4}f=\nabla_{4}\nabla_{\mu}f&=0,\quad\nabla_{4}\nabla_{4}f=f^{\prime\prime},\\
\\
\nabla_{\mu}\nabla_{\nu}f&=-e^{2A}A^{\prime}f^{\prime}\eta_{\mu\nu}.
\end{aligned}
\end{equation}
Also
\begin{equation}\label{e36.7}
\nabla^{2}f=\nabla^{C}\nabla_{C}f=(\partial_{B}\partial_{A}f-\partial_{C}f\Gamma^{C}_{\:\:\:AB})h^{AB}
=-f^{\prime\prime}-4A^{\prime}f^{\prime}.
\end{equation}
\subsection{\textsl{Minimal coupling}}
Now in this subsection we will derive the component form of the
dynamical equations for the minimal coupling of the bulk scalars
to the gravity sector and the brane forming scalar. The relative
$5D$-action is \cite{temel,tem1}
\begin{equation}\label{e37}
 S=\int\bigg(-\frac{1}{4}R_{AB}\wedge\ast e^{AB}-\ast
V-\frac{1}{4}tr(\ast d{\mathcal{M}}^{-1}\wedge
 d{\mathcal{M}})\bigg),
\end{equation}
which is differing from \eqref{e11.6} by a gravitational
coefficient. We may adopt the general results of the previous
section for the gravity-$\sigma$-model coupling bearing in mind
that now we restrict ourselves to the $5D$ geometry we have
introduced above. As usual we will assume that
$\phi^{i}=\phi^{i}(y)$ and $\chi^{\beta}=\chi^{\beta}(y)$. To
derive the component form of the energy-momentum tensor let us
first introduce
\begin{subequations}\label{e38}
\begin{align}
g_{i\theta}&=\mathcal{B}_{i\alpha}e^{\frac{1}{2}\alpha _{j}\phi
^{j}}\mathbf{\Omega}^{\alpha}_{\theta},\notag\\
g_{\tau\gamma}&=\mathcal{C}_{\alpha\beta}e^{\frac{1}{2}\alpha_{i}\phi^{i}}e^{\frac{1}{2}\beta_{j}\phi^{j}}
\mathbf{\Omega}^{\alpha}_{\tau}\mathbf{\Omega}^{\beta}_{\gamma},\notag\\
h_{i\theta\beta}&=\mathcal{B}_{i\alpha}e^{\frac{1}{2}\alpha
_{j}\phi ^{j}}\mathcal{D}^{\alpha}_{\theta\beta},\notag\\
h_{\theta\tau\gamma}&=\mathcal{C}_{\alpha\beta}e^{\frac{1}{2}\alpha_{i}\phi^{i}}e^{\frac{1}{2}\beta_{j}\phi^{j}}
\mathbf{\Omega}^{\beta}_{\gamma}\mathcal{D}^{\alpha}_{\theta\tau},\notag\\
\tag{\ref{e38}}
\end{align}
\end{subequations}
where since $\mathcal{C}_{\alpha\beta}=\mathcal{C}_{\beta\alpha}$
we have $g_{\tau\gamma}=g_{\gamma\tau}$.  In order to express the
components of the scalar field equations in a compact form in a
later subsection we also define
\begin{subequations}\label{e39}
\begin{align}
g_{ki\theta}&=\alpha_{k}\mathcal{B}_{i\alpha}e^{\frac{1}{2}\alpha
_{j}\phi ^{j}}\mathbf{\Omega}^{\alpha}_{\theta},\notag\\
\tilde{g}_{k\tau\gamma}&=(\alpha_{k}+\beta_{k})\mathcal{C}_{\alpha\beta}e^{\frac{1}{2}\alpha_{i}\phi^{i}}e^{\frac{1}{2}\beta_{j}\phi^{j}}
\mathbf{\Omega}^{\alpha}_{\tau}\mathbf{\Omega}^{\beta}_{\gamma}.
\tag{\ref{e39}}
\end{align}
\end{subequations}
With these definitions, further remembering that $D=5$,
$\mathcal{A}_{ij}=\mathcal{A}_{ji}$, and as the scalars are only
functions of $y$
$\partial^{C}(\:\cdot\:)\partial_{C}(\:\cdot\:)=-(\:\cdot\:)^{\prime}(\:\cdot\:)^{\prime}$
we can write down the components of the symmetric space sigma
model energy-momentum tensor \eqref{e25} as
\begin{subequations}\label{e40}
\begin{align}
\mathcal{T}_{44}&=-\frac{1}{8}\mathcal{A}_{ij}\phi^{i\prime}\phi^{j\prime}-\frac{1}{4}g_{i\beta}\phi^{i\prime}\chi^{\beta\prime}
-\frac{1}{2}g_{\tau\rho}\chi^{\tau\prime}\chi^{\rho\prime},\notag\\
\mathcal{T}_{\mu\nu}&=e^{2A}\eta_{\mu\nu}\big(-\frac{1}{8}\mathcal{A}_{ij}\phi^{i\prime}\phi^{j\prime}-\frac{1}{4}g_{i\beta}\phi^{i\prime}\chi^{\beta\prime}
-\frac{1}{2}g_{\tau\rho}\chi^{\tau\prime}\chi^{\rho\prime}\big),\notag\\
\mathcal{T}_{4\mu}&=\mathcal{T}_{\mu 4}=0. \tag{\ref{e40}}
\end{align}
\end{subequations}
In order to obtain the component form of the Einstein's equations
we may make use of the relation
\begin{equation}\label{e41}
 -\ast e_{ABC}\wedge R^{BC}=2G_{AB}\ast e^{B},
\end{equation}
in \eqref{e18}. Now by also performing the appropriate
normalization arising from the coefficient of the gravity sector
in \eqref{e37} we can read the component form of the Einstein's
equations from \eqref{e18} as
\begin{equation}\label{e42}
 G_{CB}=2(Vh_{CB}-\mathcal{T}_{CB}),
\end{equation}
where like in \cite{tem1} we use the convention $4\pi G=1$ for the
$5D$ bulk. By reading the appropriate metric components from
\eqref{e30}, the Einstein tensor components from \eqref{e36}, also
by using \eqref{e40} in \eqref{e42} we can derive the dynamical
equations which will shape the finite brane solutions in the $5D$
bulk. We easily see that the $G_{4\mu}$ and the $G_{\mu 4}$
components give us null results in \eqref{e42}. The non-vanishing
brane generating dynamical equations are the ones
\begin{equation}\label{e43}
\begin{aligned}
\underline{G_{\mu\nu}}\quad\quad\quad\quad\quad&\\
-3A^{\prime\prime}-6(A^{\prime})^{2}&=+2V+\frac{1}{4}\mathcal{A}_{ij}\phi^{i\prime}\phi^{j\prime}+\frac{1}{2}g_{i\beta}\phi^{i\prime}\chi^{\beta\prime}
+g_{\tau\rho}\chi^{\tau\prime}\chi^{\rho\prime},\\\\
\underline{G_{44}}\quad\quad\quad\quad\quad&\\
6(A^{\prime})^{2}&=-2V+\frac{1}{4}\mathcal{A}_{ij}\phi^{i\prime}\phi^{j\prime}+\frac{1}{2}g_{i\beta}\phi^{i\prime}\chi^{\beta\prime}
+g_{\tau\rho}\chi^{\tau\prime}\chi^{\rho\prime}.
\end{aligned}
\end{equation}
These two equations can be combined to yield
\begin{equation}\label{e44}
 A^{\prime\prime}=-\frac{1}{6}\mathcal{A}_{ij}\phi^{i\prime}\phi^{j\prime}-\frac{1}{3}g_{i\beta}\phi^{i\prime}\chi^{\beta\prime}
-\frac{2}{3}g_{\tau\rho}\chi^{\tau\prime}\chi^{\rho\prime},
\end{equation}
which together with the second equation in \eqref{e43} is more
appropriate for the first-order formalism in search for brane
solutions. On the other hand as discussed in \cite{temel} one can
show that solutions of the first-order equations
\begin{equation}\label{e44.5}
 \varphi^{a\,\prime}=\frac{1}{2}g^{ab}\partial_{\varphi^{b}}W(\varphi^{c}),\quad
 A^{\prime}=-\frac{1}{3}W(\varphi^{c}),
\end{equation}
are also solutions of \eqref{e43},\eqref{e44} and the
corresponding scalar field equations. Here supergravity originated
superpotential $W$ is related to the ordinary potential $V$ via
\begin{equation}\label{e44.6}
 V(\varphi^{c})=\frac{1}{8}g^{ab}\partial_{\varphi^{a}}W(\varphi^{c})\partial_{\varphi^{b}}W(\varphi^{c})-\frac{1}{3}\big(W(\varphi^{c})\big)^{2}.
\end{equation}
In this expression without sub-labelling we have denoted the set
of scalars $\phi^{i}$ and $\chi^{\beta}$ generally as
$\varphi^{c}$. The scalar manifold metric $g_{ab}$ which is
introduced in \eqref{e1} can be directly read from \eqref{e6}
which is also explicitly studied in \cite{branemot}.
\subsection{\textsl{Non-minimal coupling}}
In this subsection we will consider the generalization of the
action \eqref{e37} in the gravity-scalar coupling sense and we
will derive the dynamical field equations of the non-minimal
coupling of the scalars to the gravity. The action which contains
the modified gravity-bulk scalar coupling takes the form
\begin{equation}\label{e45}
 S=\int\bigg(-\frac{1}{4}\textit{f}\:R_{AB}\wedge\ast e^{AB}-\ast
V-\frac{1}{4}tr(\ast d{\mathcal{M}}^{-1}\wedge
 d{\mathcal{M}})\bigg),
\end{equation}
where $\textit{f}=\textit{f}(\phi^{i},\chi^{\beta})$ is a generic
function of the bulk scalars. In the Appendix we present the
details of the variation of the gravity term in \eqref{e45}. By
using \eqref{app16} \footnote{By also bearing in mind the
appropriate normalization for the energy-momentum tensor
\eqref{e24.5} which relates it to the component-form variation of
\eqref{app16}.} via the total variation of \eqref{e45} we can
write down the modified Einstein's equations as
\begin{equation}\label{e46}
-\frac{1}{2}\textit{f}\:G_{AB}+\frac{1}{2}\big(\nabla_{A}\nabla_{B}(\:\textit{f}\:)-\nabla_{C}\nabla^{C}(\:\textit{f}\:)h_{AB}\big)
 =-Vh_{AB}+\mathcal{T}_{AB}.
\end{equation}
Now if we consider the $5D$ manifold with the metric \eqref{e30}
since $G_{4\nu};\:$$\nabla_{4}\nabla_{\nu}(\:\textit{f}\:),$ as
well as $h_{4\nu},\:$$\mathcal{T}_{4\nu}$ vanish \eqref{e46} gives
us null equations for the $G_{4\nu}$ and $G_{\nu4}$ components.
Similarly by substituting the other components via \eqref{e30},
\eqref{e36}, \eqref{e36.6}, \eqref{e40} in \eqref{e46} we get the
non-vanishing finite-brane generating dynamical equations of the
non-minimal gravity coupling of the bulk scalars as
\begin{equation}\label{e47}
\begin{aligned}
\underline{G_{\mu\nu}}\quad\quad\quad\quad\quad\quad\quad\quad\quad&\\
3\textit{f}\:\big(A^{\prime\prime}+2(A^{\prime})^{2}\big)+\textit{f}^{\:\prime\prime}+3\:\textit{f}^{\:\prime}A^{\prime}
&=-2V-\frac{1}{4}\mathcal{A}_{ij}\phi^{i\prime}\phi^{j\prime}-\frac{1}{2}g_{i\beta}\phi^{i\prime}\chi^{\beta\prime}
-g_{\tau\rho}\chi^{\tau\prime}\chi^{\rho\prime},\\\\
\underline{G_{44}}\quad\quad\quad\quad\quad\quad\quad\quad\quad&\\
-6\textit{f}\:(A^{\prime})^{2}-4\:\textit{f}^{\:\prime}A^{\prime}&=2V-\frac{1}{4}\mathcal{A}_{ij}\phi^{i\prime}\phi^{j\prime}-\frac{1}{2}g_{i\beta}\phi^{i\prime}\chi^{\beta\prime}
-g_{\tau\rho}\chi^{\tau\prime}\chi^{\rho\prime}.
\end{aligned}
\end{equation}
By again using the second one in the first we obtain
\begin{equation}\label{e48}
 3\textit{f}\:A^{\prime\prime}+\textit{f}^{\:\prime\prime}-\textit{f}^{\:\prime}A^{\prime}=-\frac{1}{2}\mathcal{A}_{ij}\phi^{i\prime}\phi^{j\prime}
 -g_{i\beta}\phi^{i\prime}\chi^{\beta\prime}
-2g_{\tau\rho}\chi^{\tau\prime}\chi^{\rho\prime},
\end{equation}
which again together with $G_{44}$-component equation in
\eqref{e47} is more appropriate for a first-order formulation to
search for finite brane solutions. We should state that likewise
the minimal coupling case one can also work on the construction of
an equivalent first-order formalism in terms of a superpotential
$W$.
\subsection{\textsl{Scalar field equations}}
We now consider the scalar sector and we will present the
component form of the scalar field equations for both of the
couplings discussed above. Beforehand to be able to switch our
formulation to the component form we should remind the reader of
the identity
\begin{equation}\label{e49}
e^{A}\wedge \ast e^{B}=h^{AB}\ast 1,
\end{equation}
which is valid for a generic moving co-frame. Furthermore for a
one-form $A=A_{C}e^{C}$ and a scalar field $\varphi$ we have
\begin{equation}\label{e50}
\begin{aligned}
d\ast A=\big(\nabla_{C}A^{C}\big)\ast 1&=\big(\partial_{C}A^{C}+A^{C}\Gamma^{B}_{\:\:\:CB}\big)\ast 1, \\\\
d\ast d\varphi&=\big(\nabla^{2}\varphi\big)\ast 1,
\end{aligned}
\end{equation}
where $\nabla^{2}$ is defined in \eqref{e36.7}. Now for the
definitions in \eqref{e38} let us introduce the coefficients
\begin{equation}\label{e51}
\begin{aligned}
dg_{k\beta}&=X_{ik\beta}d\phi^{i}+X_{\theta
k\beta}d\chi^{\theta},\\\\
dg_{\gamma\theta}&=X_{i\gamma\theta}d\phi^{i}+X_{\kappa\gamma\theta}d\chi^{\kappa},
\end{aligned}
\end{equation}
where we define
\begin{equation}\label{e51.5}
\begin{aligned}
X_{ik\beta}&=\frac{\partial g_{k\beta}}{\partial\phi^{i}},\quad
X_{\theta
k\beta}&=\frac{\partial g_{k\beta}}{\partial\chi^{\theta}},\\\\
X_{i\gamma\theta}&=\frac{\partial
g_{\gamma\theta}}{\partial\phi^{i}},\quad
X_{\kappa\gamma\theta}&=\frac{\partial
g_{\gamma\theta}}{\partial\chi^{\kappa}}.
\end{aligned}
\end{equation}
Following the definitions introduced in \eqref{e38} and
furthermore by making use of the above definitions, from
\eqref{e15.5} and \eqref{e15.6} we can express the scalar field
equations in a more compact form as
\begin{subequations}\label{e52}
\begin{align}
\mathcal{A}_{ik}d\ast d\phi^{i}+g_{k\beta}d \ast
d\chi^{\beta}&=\big(\frac{1}{2}g_{ki\beta}-X_{ik\beta}\big)d\phi^{i}\wedge\ast
d \chi^{\beta}\notag\\
&\:\:\:+\big(\tilde{g}_{k\tau\gamma}-X_{\gamma k
\tau}\big)d\chi^{\gamma}\wedge\ast d \chi^{\tau}
+F_{k}(V),\notag\\\notag\\
\frac{1}{2}g_{i\theta}d\ast d\phi^{i}+2g_{\theta\gamma}d \ast
d\chi^{\gamma}&=\big(\frac{1}{2}h_{i\theta\beta}-\frac{1}{2}X_{\beta
i\theta}-2X_{i\theta\beta}\big)d\phi^{i}\wedge\ast d
\chi^{\beta}\notag\\
&\:\:\:+\big(h_{\theta\tau\gamma}+h_{\theta\gamma\tau}-2X_{\gamma\theta\tau}\big)d\chi^{\gamma}\wedge\ast
d \chi^{\tau}\notag\\
&\:\:\:-\frac{1}{2}X_{ji\theta}d\phi^{j}\wedge\ast d
\phi^{i}+F_{\theta}(V), \tag{\ref{e52}}
\end{align}
\end{subequations}
where we have introduced
\begin{equation}\label{e53}
\begin{aligned}
\underline{\textit{Minimal coupling}}\:\quad\quad\quad&\\
F_{k}(V)&=-4\partial_{\phi^{k}}V\ast 1\equiv\mathcal{F}_{k}(V)\ast 1,\\\\
F_{\theta}(V)&=-2\partial_{\chi^{\theta}}V\ast
1\equiv\mathcal{F}_{\theta}(V)\ast 1,\\\\
\underline{\textit{Non-minimal coupling}}\:\quad\quad\quad&\\
F_{k}(V)&=\big(-4\partial_{\phi^{k}}V-\frac{1}{4}\big(\partial_{\phi^{k}}\:\textit{f}\:\big)R\big)\ast 1\equiv\mathcal{F}_{k}(V)\ast 1,\\\\
F_{\theta}(V)&=\big(-2\partial_{\chi^{\theta}}V
-\frac{1}{4}\big(\partial_{\chi^{\theta}}\:\textit{f}\:\big)R\big)\ast
1\equiv\mathcal{F}_{\theta}(V)\ast 1.
\end{aligned}
\end{equation}
The $F$-terms include the contributions to the scalar sector
coming from the variation of the potential term for both of the
cases whereas there is an extra contribution for the non-minimal
coupling case which arises from the variation of the modified
Hilbert-Einstein term in the action. Now by using \eqref{e49} and
\eqref{e50} we can first express the component form of the scalar
field equations \eqref{e52} then since we assume that
$\phi^{i}=\phi^{i}(y)$ and $\chi^{\beta}=\chi^{\beta}(y)$ via the
relations in \eqref{e36.6} and \eqref{e36.7} we can obtain the
scalar field equations for the $5D$-bulk geometry arising from the
metric \eqref{e30}. The result reads
\begin{equation}\label{e54}
\begin{aligned}
\underline{\textit{Dilaton equations}}\quad\quad\quad\quad\quad\quad&\\\\
\mathcal{A}_{ik}\big(-\phi^{i\,\prime\:\prime}-4A^{\prime}\phi^{i\,\prime}\big)+g_{k\beta}\big(&-\chi^{\beta\,\prime\,\prime}-4A^{\prime}\chi^{\beta
\,\prime}\big)\\
&=-\big(\frac{1}{2}g_{ki\beta}-X_{ik\beta}\big)\phi^{i\,\prime}
\chi^{\beta\,\prime}\\
&\:\:\:\:\:-\big(\tilde{g}_{k\tau\gamma}-X_{\gamma k
\tau}\big)\chi^{\gamma\,\prime}\chi^{\tau\,\prime}
+\mathcal{F}_{k}(V),\\
\underline{\textit{Axion equations}}\quad\quad\quad\quad\quad\quad&\\\\
\frac{1}{2}g_{i\theta}\big(-\phi^{i\,\prime\,\prime}-4A^{\prime}\phi^{i\,\prime}\big)+2g_{\theta\gamma}
\big(&-\chi^{\gamma\,\prime\,\prime}-4A^{\prime}\chi^{\gamma\,\prime}\big)\\
&=-\big(\frac{1}{2}h_{i\theta\beta}-\frac{1}{2}X_{\beta
i\theta}-2X_{i\theta\beta}\big)\phi^{i\,\prime}
\chi^{\beta\,\prime}\\
&\:\:\:\:\:-\big(h_{\theta\tau\gamma}+h_{\theta\gamma\tau}-2X_{\gamma\theta\tau}\big)\chi^{\gamma\,\prime}
\chi^{\tau\,\prime}\\
&\:\:\:\:\:+\frac{1}{2}X_{ji\theta}\phi^{j\,\prime}
\phi^{i\,\prime}+\mathcal{F}_{\theta}(V).
\end{aligned}
\end{equation}
These equations are the coupled scalar equations for the brane
forming scalar and the bulk scalars which must be solved
simultaneously with \eqref{e43}, \eqref{e44} for the minimal
coupling and with \eqref{e47},\eqref{e48} for the non-minimal
coupling cases.
\section{Conclusion}
In Section two, we have discussed the elements of the general
action which couples gravity to the symmetric space sigma model in
the solvable Lie algebra parametrization of the symmetric target
space. We have performed the variation of the action to obtain the
corresponding Einstein's equations. The variation also leads to
the scalar field equations which coincide with the ones derived in
\cite{sssm1} which are obtained by direct application of the
Euler-Lagrange equations. After identifying the energy-momentum
one-forms from the Einstein's equations we have calculated the
energy-momentum tensor associated with a generic symmetric space
sigma model action. Then in Section three we have turned our
attention to the thick brane scenarios of fundamental
interactions. Following the discussion about the constituents of
the warped $5D$ bulk geometry which accommodates finite thick
brane solutions we have focussed on the dynamics of the bulk
scalars in this framework. In this respect we have used the
results of the previous section to derive the dynamical equations
for the models of thick branes in which the bulk scalar
interactions are assumed to be localized on the finitely sized
brane solution that is formed by one of the scalars. As we have
discussed before such a model is described by a
$\sigma$-model-gravity coupling in the context of warped bulk
geometry. As we have mentioned in the Introduction our focus has
been on the interactions generated by symmetric space sigma models
which appear as the most common scalar sectors in the dimensional
reduction of supergravities. Our analysis has included two
distinct cases of minimal and non-minimal scalar-gravity couplings
and for both of these cases we have derived the component form of
the dynamical equations appropriate for a first-order formalism.
Finally we also obtained the scalar field equations for both of
the cases in component form for the warped bulk geometry.

Although our analysis is performed for a general $\sigma$-model
coupling it is done for the flat-brane case. One may consider
extensions of the interaction formalism presented here to the
$AdS$ or $dS$ brane geometries to study localized gravity in these
models as well as the corresponding dual $CFT$ renormalization
group flow equations \cite{tem2}. Similarly one may also consider
cosmological solutions in the $4D$-sector including scalar
interactions. Other localized interactions of the bulk
(specifically for various supergravity multiplets) can also be
considered in connection with the scalar sector which is studied
in detail here as it needs special care owing to its non-linear
sigma model structure. We should state that the formulation
presented here is purely algebraic and general, it is applicable
to the sigma models whose target spaces are generic symmetric
spaces of the form $G/K$. Therefore the results provide valuable
and case-free tools for any other coupling extensions. From this
point of view our formulation presents a formal framework for the
study of specifically chosen localized supergravity theories in
the context of thick brane scenarios. In this work although we
have derived the necessary dynamical field equations for a class
of models we have not attempted to construct the first-order
formalisms and studied their solutions. Starting from the
currently occurring Grasmannian scalar manifold models and
supersymmetry predicted superpotentials of supergravity theories
one may work out for either generic multiplets or for specific
supergravity theories interaction carrying extended thick brane
solutions in parallel with the rich literature on non-interacting
single scalar models. One may also study bulk scalar interactions
on the brane for higher dimensional bulk geometries in connection
with higher dimensional supergravities. We should also remark that
similar analysis can also be extended for generalized scalar and
gravity dynamics \cite{tem1,baz10}. Finally apart from the
specific directions we have mentioned above the results of this
work can find extensions in various aspects of thick brane
dynamics including search for particular solutions and probing
phenomenology of localized gravity, gauge and matter fields in
various geometries
\cite{410,baz1,baz2,baz3,baz4,baz5,baz6,baz7,baz8}.

In relevance to the mathematical content of the present work one
may also slightly change the point of view from the
warped-geometric braneworld cosmological scenario to the
consideration of the coupling of Section two on its own right.
That is to say the sigma model can purely be considered as a
non-linear source to the gravity sector. It can separately be used
as an ingredient in the construction of UV completion of GR in a
non-linear context. On the gauge theory side in the UV completion
by spontaneous symmetry breaking of the low energy effective sigma
model theory which is also closely related to the string theories
sigma models play the central role. One may expect a similar
contribution to the quantum gravity model building which also
emerges from the unifying and underlying string theories. Recently
a ghost-free non-linear extension of the Fierz-Pauli \cite{fp}
massive gravity has been constructed
\cite{ah,dgrt1,dgrt2,crem,hr1,hr2}. In its physically truncated
form \cite{hr1} of this non-linear massive gravity the basic
ingredient of the sigma model kinetic term namely the induced
metric \eqref{e3} appears as an argument of the graviton potential
which gives the physical metric its mass as a result of a
gravitational Higgs mechanism. In that formalism \eqref{e3}
becomes the so-called fiducial metric $f_{\mu\nu}$; the basic
field which is a sigma model type kinetic term and which enables
the coupling of the Sk\"{u}ckelberg scalars to the physical
metric. This emergence suggests that sigma models may play a more
generalized role with their entire geometrical content in massive
as well as massless non-linear extensions of GR. We should share
an observation here that introducing sigma models in massive
graviton constructions would need to do changes in the link
structure between the sites of the theory space construction of
\cite{ah} which inspired the de Rham, Gabadadze, Tolley (dRGT)
programme of massive gravity. In \cite{ah} the links were taken as
general coordinate transformations. To introduce sigma models one
should also impose the embedding structure of them onto the links,
possibly one may introduce immersion maps between various sites in
the theory space and consider objects which not only transform
covariently under both of the general coordinate transformations
via links but also respect to the immersion structure. This way of
introducing sigma models at link level may also resemble the
situation valid at low energies for certain gauge theory
constructions.

Apart from this above-mentioned fundamental role, non-linear sigma
models can also be used as a direct non-linear source in
non-linear massive gravity. This can possibly be done in three
ways. Firstly in \cite{ah} the dual site on which the fiducial
metric lives is also taken to be $4D$ so that the links namely the
St\"{u}ckelberg fields which can be considered as coordinate
transformations have correct number of dofs (for which one is
identified as the Boulware-Deser (BD) ghost to be erased). One may
introduce more than four scalars in the context of sigma model
immersion framework for which only four of them will get a vev via
$\phi^{\mu}=x^{\mu}+\delta\phi^{\mu}$ where
$\delta\phi^{\mu}\equiv\pi^{\mu}$ are the Goldstone bosons which
represent small fluctuations around a background spacetime. Then
the relevance of the rest of the scalars can be inspected in the
modification of GR to explain DM, DE, or inflationary structures
arising via non-linear level interactions of these extra
geometrical dof's. Secondly, in \cite{aco} it has been shown that
in the non-linear massive gravity dRGT theory when one chooses the
fiducial metric flat the homogeneous and isotropic FRW type
cosmological solutions do not exist due to the same mechanism
which cancels the BD ghost. Recently the de Sitter (which is a
symmetric space) choice of the fiducial metric is being studied
(see for example \cite{tolley,ln}). These constructions admit FRW
type solutions showing once more the special place of de Sitter
space in massive gravity theories but still they posses some
physical problems to be resolved. At this point in connection with
the formalism presented in this work and in \cite{branemot} one
may introduce a parametric generalization for the de Sitter case
namely taking the fiducial metric as the most general form of the
induced metric of the symmetric space sigma model may assist in
the search for the true fiducial background to obtain a physical
homogeneous and isotropic universe model from the massive gravity
theory. Finally as a third interface between the non-linear
massive gravity and the non-linear sigma models like in the
attempts to introduce dynamics to the fiducial metric
\cite{has1,has2,comel,crisos,khosra,nomura} one may take the
fiducial metric as the induced metric on the immersed p-brane and
consider p-brane massive gravity interactions as a new non-linear
source to the dRGT massive gravity.
\section{Appendix}
\appendix
\section{Variational Details}
Here we present the details of the variation of the
gravity-$\sigma$-model coupling of Section two and the non-minimal
gravity-scalar coupling of Section three.
\subsection{SSSM-gravity coupling}
In spite of the fact that the results coincide with the standard
variational methods here we will present a detailed account of the
global variation of the total action in \eqref{e12}. When one
applies a variation operator on \eqref{e12} one can consider the
variation of the orthogonal moving co-frame $\{e^{A}\}$; $\delta
e^{C}$ which contains the variation of the metric $h_{AB}$ in it
and the variation of the scalar fields namely $\delta \phi^{j},
\delta \chi^{\tau}$ separately. Therefore one may collect the
relative terms in two disjoint groups as in \eqref{e13}. The first
term which contains the gravitational variation becomes
\begin{subequations}\label{app1}
\begin{align}
\delta S_{1}=\int\bigg(&(-1)^{D-2}d(\ast
e^{AB}\wedge\delta\omega_{AB})+\delta e^{C}\wedge\bigg(\ast
e_{ABC}\wedge R^{AB}-V\ast
e_{C}\notag\\
&-(-1)^{D-1}\frac{1}{8}\,\mathcal{A}_{ij}(d{\phi}^{i}\wedge i_{C}
\ast d\phi^{j} +i_{C}d{\phi}^{j}\wedge\ast
d\phi^{i})-(-1)^{D-1}\frac{1}{4}\,\mathcal{B}_{i\alpha}\notag\\
&\times e^{\frac{1}{2}\alpha_{j}\phi^{j}}
\mathbf{\Omega}^{\alpha}_{\beta}(d{\phi}^{i}\wedge i_{C} \ast
d\chi^{\beta} +i_{C}d\chi^{\beta}\wedge\ast d\phi^{i})
-(-1)^{D-1}\frac{1}{2}\,\mathcal{C}_{\alpha\beta}\notag\\
&\times
e^{\frac{1}{2}\alpha_{i}\phi^{i}}e^{\frac{1}{2}\beta_{j}\phi^{j}}
\mathbf{\Omega}^{\alpha}_{\tau}\mathbf{\Omega}^{\beta}_{\rho}(d\chi^{\tau}\wedge
i_{C} \ast d\chi^{\rho} +i_{C}d\chi^{\rho}\wedge\ast
d\chi^{\tau})\bigg)\bigg),\notag\\
\tag{\ref{app1}}
\end{align}
\end{subequations}
and the second term which is composed of the scalar field
variations reads
\begin{subequations}\label{app2}
\begin{align}
\delta
S_{2}=\int\bigg(&(-1)^{D-1}\frac{1}{8}\,\mathcal{A}_{ij}(\delta
d{\phi}^{j}\wedge\ast d\phi^{i} +\delta d{\phi}^{i}\wedge\ast
d\phi^{j})\notag\\
&+(-1)^{D-1}\frac{1}{4}\,\mathcal{B}_{i\alpha}
\bigg(\delta(e^{\frac{1}{2}\alpha_{j}\phi^{j}}
\mathbf{\Omega}^{\alpha}_{\beta})d\chi^{\beta}\wedge\ast
d\phi^{i}+e^{\frac{1}{2}\alpha_{j}\phi^{j}}
\mathbf{\Omega}^{\alpha}_{\beta}\notag\\
&\times(\delta d\chi^{\beta}\wedge \ast d\phi^{i} +\delta
d\phi^{i}\wedge\ast d\chi^{\beta})\bigg)
+(-1)^{D-1}\frac{1}{2}\,\mathcal{C}_{\alpha\beta}\notag\\
&\times\bigg(
\delta(e^{\frac{1}{2}\alpha_{i}\phi^{i}}e^{\frac{1}{2}\beta_{j}\phi^{j}}
\mathbf{\Omega}^{\alpha}_{\tau}\mathbf{\Omega}^{\beta}_{\gamma})
d\chi^{\gamma}\wedge \ast d\chi^{\tau}
+e^{\frac{1}{2}\alpha_{i}\phi^{i}}e^{\frac{1}{2}\beta_{j}\phi^{j}}
\mathbf{\Omega}^{\alpha}_{\tau}\mathbf{\Omega}^{\beta}_{\gamma}\notag\\
&\times(\delta d\chi^{\gamma}\wedge\ast d\chi^{\tau} +\delta
d\chi^{\tau}\wedge\ast
d\chi^{\gamma})\bigg)-\bigg(\partial_{i}V\delta\phi^{i}+\partial_{\alpha}V\delta\chi^{\alpha}\bigg)\ast
1\bigg), \tag{\ref{app2}}
\end{align}
\end{subequations}
where we define $\partial_{i}V\equiv\frac{\partial
V}{\partial\phi^{i}}$ and  $\partial_{\alpha}V\equiv\frac{\partial
V}{\partial\chi^{\alpha}}$. One obtains the Einstein's equations
\eqref{e18} by equating the coefficients of the variations of the
co-frame to zero in \eqref{app1}. As usual the first term in
\eqref{app1} which is an exact differential form gives a null
result in the variation due to the standard assumptions of the
variational method \cite{thring} thus it does not contribute to
the Einstein's equations. In general when $N$ is an r-chain one
can apply the Stoke's theorem and then one can assume that the
variation of the independent fields vanish on the boundary
$\partial N$ of $N$. If $\partial N=\emptyset$ then directly the
integral of the first term vanishes. We can furthermore simplify
\eqref{app2} to write it in its final form in which we gather the
coefficients of the variations $\delta\phi^{i}$ and
$\delta\chi^{\theta}$. The result becomes
\begin{subequations}\label{app3}
\begin{align}
\delta
S_{2}=\int\bigg(&(-1)^{D-1}\frac{1}{8}\,\mathcal{A}_{ij}d(\delta
\phi^{j}\wedge\ast
d\phi^{i})+(-1)^{D-1}\frac{1}{8}\,\mathcal{A}_{ij}d(\delta\phi^{i}\wedge\ast
d\phi^{j})\notag\\
&+(-1)^{D-1}\frac{1}{4}\,\mathcal{B}_{i\alpha}
d(\delta\chi^{\beta}\wedge e^{\frac{1}{2}\alpha_{j}\phi^{j}}
\mathbf{\Omega}^{\alpha}_{\beta}\ast d\phi^{i})\notag\\
&+(-1)^{D-1}\frac{1}{4}\,\mathcal{B}_{i\alpha}
d(\delta\phi^{i}\wedge e^{\frac{1}{2}\alpha_{j}\phi^{j}}
\mathbf{\Omega}^{\alpha}_{\beta}\ast d\chi^{\beta})\notag\\
&+(-1)^{D-1}\frac{1}{2}\,\mathcal{C}_{\alpha\beta} d(\delta
\chi^{\gamma}\wedge
e^{\frac{1}{2}\alpha_{i}\phi^{i}}e^{\frac{1}{2}\beta_{j}\phi^{j}}
\mathbf{\Omega}^{\alpha}_{\tau}\mathbf{\Omega}^{\beta}_{\gamma}\ast
d\chi^{\tau})\notag\\
&+(-1)^{D-1}\frac{1}{2}\,\mathcal{C}_{\alpha\beta} d(\delta
\chi^{\tau}\wedge
e^{\frac{1}{2}\alpha_{i}\phi^{i}}e^{\frac{1}{2}\beta_{j}\phi^{j}}
\mathbf{\Omega}^{\alpha}_{\tau}\mathbf{\Omega}^{\beta}_{\gamma}\ast
d\chi^{\gamma})\notag\\
&+\delta\phi^{n}\bigg(-\partial_{n}V\ast
1-(-1)^{D-1}\frac{1}{8}\,\mathcal{A}_{in}d\ast
d\phi^{i}-(-1)^{D-1}\frac{1}{8}\,\mathcal{A}_{nj}d\ast
d\phi^{j}\notag\\
&+(-1)^{D-1}\frac{1}{8}\,\mathcal{B}_{i\alpha}
\alpha_{n}e^{\frac{1}{2}\alpha_{j}\phi^{j}}
\mathbf{\Omega}^{\alpha}_{\beta}d\chi^{\beta}\wedge\ast
d\phi^{i}\notag\\
&-(-1)^{D-1}\frac{1}{4}\,\mathcal{B}_{n\alpha}
d(e^{\frac{1}{2}\alpha_{j}\phi^{j}}
\mathbf{\Omega}^{\alpha}_{\beta}\ast d\chi^{\beta})
+(-1)^{D-1}\frac{1}{2}\,\mathcal{C}_{\alpha\beta}\notag\\
&\times(
\frac{1}{2}\alpha_{n}e^{\frac{1}{2}\alpha_{i}\phi^{i}}e^{\frac{1}{2}\beta_{j}\phi^{j}}
+\frac{1}{2}\beta_{n}e^{\frac{1}{2}\alpha_{i}\phi^{i}}e^{\frac{1}{2}\beta_{j}\phi^{j}})
\mathbf{\Omega}^{\alpha}_{\tau}\mathbf{\Omega}^{\beta}_{\gamma}
d\chi^{\gamma}\wedge \ast d\chi^{\tau}\bigg)\notag\\
&+\delta\chi^{\theta}\bigg(-\partial_{\theta}V\ast
1+(-1)^{D-1}\frac{1}{4}\,\mathcal{B}_{i\alpha}
e^{\frac{1}{2}\alpha_{j}\phi^{j}}
\mathcal{D}^{\alpha}_{\theta\beta}d\chi^{\beta}\wedge\ast
d\phi^{i}\notag\\
&-(-1)^{D-1}\frac{1}{4}\,\mathcal{B}_{i\alpha}
d(e^{\frac{1}{2}\alpha_{j}\phi^{j}}
\mathbf{\Omega}^{\alpha}_{\theta}\ast d\phi^{i})+(-1)^{D-1}\frac{1}{2}\,\mathcal{C}_{\alpha\beta}\notag\\
&\times(\mathcal{D}^{\alpha}_{\theta\tau}\mathbf{\Omega}^{\beta}_{\gamma}+\mathbf{\Omega}^{\alpha}_{\tau}
\mathcal{D}^{\beta}_{\theta\gamma})e^{\frac{1}{2}\alpha_{i}\phi^{i}}e^{\frac{1}{2}\beta_{j}\phi^{j}}
d\chi^{\gamma}\wedge \ast d\chi^{\tau}\notag\\
&-(-1)^{D-1}\frac{1}{2}\,\mathcal{C}_{\alpha\beta} d(
e^{\frac{1}{2}\alpha_{i}\phi^{i}}e^{\frac{1}{2}\beta_{j}\phi^{j}}
\mathbf{\Omega}^{\alpha}_{\tau}\mathbf{\Omega}^{\beta}_{\theta}\ast
d\chi^{\tau})\notag\\
&-(-1)^{D-1}\frac{1}{2}\,\mathcal{C}_{\alpha\beta} d(
e^{\frac{1}{2}\alpha_{i}\phi^{i}}e^{\frac{1}{2}\beta_{j}\phi^{j}}
\mathbf{\Omega}^{\alpha}_{\theta}\mathbf{\Omega}^{\beta}_{\gamma}\ast
d\chi^{\gamma})\bigg)\bigg), \tag{\ref{app3}}
\end{align}
\end{subequations}
where we have introduced the matrix functions
\begin{equation}\label{app4}
\mathcal{D}_{\theta}\equiv\frac{\partial\mathbf{\Omega}}{\partial\chi^{\theta}}.
\end{equation}
As we have pointed out earlier we do not need their explicit form
in our formulation in this work. The reader may refer to their
formal derivation and therein their exact definitions in
\cite{sssm1}. If we equate the coefficients of $\delta\phi^{n}$
and $\delta\chi^{\theta}$ to zero, also by again disregarding the
exact-forms in \eqref{app3} which give surface terms we may obtain
the field equations for the scalars \eqref{e15.5} and
\eqref{e15.6}.  These field equations are the ones which are
already derived in \cite{sssm1}.
\subsection{Non-minimal gravity-scalar
coupling}
 Here we will take a look at the variation of the gravitational
term in \eqref{e45} which is the differing term from \eqref{e37}.
We will derive the variation for a generic $D$-dimensional
manifold $N$. By applying a standard analysis we can write
\begin{subequations}\label{app5}
\begin{align}
\delta\big(-\frac{1}{4}\int\textit{f}\:\ast
R\big)&=-\frac{1}{4}\int\delta\textit{f}\:\ast R
-\frac{1}{4}\int\textit{f}\:\delta(\ast R)\notag\\
&=-\frac{1}{4}\int\big(\partial_{i}\textit{f}\:\delta\phi^{i}+\partial_{\alpha}\textit{f}\:\delta\chi^{\alpha}\big)\ast
R -\frac{1}{4}\int\textit{f}\:G_{AB}\delta h^{AB}\ast 1\notag\\
&\:\:\:\:\:-\frac{1}{4}\int\textit{f}\:\delta R_{AB}h^{AB}\ast 1.
\tag{\ref{app5}}
\end{align}
\end{subequations}
For a coordinate moving co-frame starting from the definition of
the Ricci tensor via \eqref{e34} and by using the metric
compatibility
\begin{equation}\label{app6}
\nabla_{C}h^{AB}=0,
\end{equation}
after some algebra one can show that
\begin{equation}\label{app7}
-\frac{1}{4}\int\textit{f}\:\delta R_{AB}h^{AB}\ast
1=-\frac{1}{4}\int\textit{f}\:\nabla_{C}K^{C}\ast 1,
\end{equation}
where
\begin{equation}\label{app8}
K^{C}=h^{AB}\delta\Gamma^{C}_{\:\:\:BA}-h^{AC}\delta\Gamma^{B}_{\:\:\:BA}.
\end{equation}
Furthermore we have
\begin{equation}\label{app9}
-\frac{1}{4}\int\textit{f}\:\nabla_{C}K^{C}\ast
1=-\frac{1}{4}\int\nabla_{C}(\textit{f}\:K^{C})\ast
1+\frac{1}{4}\int\nabla_{C}(\:\textit{f}\:)K^{C}\ast 1.
\end{equation}
By using the metric compatibility \eqref{app6} once more we can
show that the second term in \eqref{app9} can be written as
\begin{subequations}\label{app10}
\begin{align}
\frac{1}{4}\int\nabla_{C}(\:\textit{f}\:)K^{C}\ast
1&=-\frac{1}{4}\int\nabla_{C}(\:\textit{f}\:)\nabla_{B}\delta
h^{CB}\ast
1\notag\\
&\:\:\:\:\:-\frac{1}{2}\int\nabla_{C}(\:\textit{f}\:)\delta\Gamma^{B}_{\:\:\:AB}h^{CA}\ast
1.\tag{\ref{app10}}
\end{align}
\end{subequations}
Now the first term on the right hand side of this equation can
further be expressed as
\begin{subequations}\label{app11}
\begin{align}
-\frac{1}{4}\int\nabla_{C}(\:\textit{f}\:)\nabla_{B}\delta
h^{CB}\ast
1&=-\frac{1}{4}\int\nabla_{B}\big(\nabla_{C}(\:\textit{f}\:)\delta
h^{CB}\big)\ast
1\notag\\
&\:\:\:\:\:+\frac{1}{4}\int\nabla_{B}\big(\nabla_{C}(\:\textit{f}\:)\big)\delta
h^{CB}\ast 1.\tag{\ref{app11}}
\end{align}
\end{subequations}
By using the volume form via
\begin{equation}\label{app12}
\ast 1=dx^{D}\sqrt{|deth|},
\end{equation}
and the identity
\begin{equation}\label{app13}
N^{C}\delta\Gamma^{B}_{\:\:\:CB}=\delta(\nabla_{C}N^{C})-\nabla_{C}(\delta
N^{C}),
\end{equation}
where
\begin{equation}\label{app13.5}
N^{C}=\nabla_{A}(\:\textit{f}\:)h^{AC},
\end{equation}
after some algebra by also using the Stoke's theorem we can
express the second term on the right hand side of \eqref{app10} as
\begin{subequations}\label{app14}
\begin{align}
-\frac{1}{2}\int\nabla_{C}(\:\textit{f}\:)\delta\Gamma^{B}_{\:\:\:AB}h^{CA}\ast
1&=-\frac{1}{2}\underset{\partial
N}{\int}dx^{(D-1)}\delta(\:\sqrt{|det\tilde{h}|}\:)N^{C}n_{C}\notag\\
&\:\:\:\:\:+\frac{1}{2}\int
dx^{D}\delta(\:\sqrt{|deth|}\:)\nabla_{C}N^{C},\tag{\ref{app14}}
\end{align}
\end{subequations}
where $\{n_{C}\}$ is the unit normal vector to the boundary
$\partial N$ of $N$ and in the first term on the right hand side
$\tilde{h}$ must be taken as the image of the $D$-dimensional
metric $h$ under the inclusion map of the boundary $\partial N$ in
the manifold $N$. Since
\begin{equation}\label{app15}
\delta(\:\sqrt{|deth|}\:)=-\frac{1}{2}\sqrt{|deth|}h_{AB}\delta
h^{AB},
\end{equation}
we can finally write
\begin{subequations}\label{app16}
\begin{align}
\delta\big(-\frac{1}{4}\int\textit{f}\:\ast R\big)
&=-\frac{1}{4}\int\big(\partial_{i}\textit{f}\:\delta\phi^{i}+\partial_{\alpha}\textit{f}\:\delta\chi^{\alpha}\big)\ast
R -\frac{1}{4}\int\textit{f}\:G_{AB}\delta h^{AB}\ast 1\notag\\
&\:\:\:\:\:+\frac{1}{4}\int\nabla_{B}\big(\nabla_{C}(\:\textit{f}\:)\big)\delta
h^{CB}\ast
1\notag\\
&\:\:\:\:\:-\frac{1}{4}\int\nabla_{C}\big(\nabla_{A}(\:\textit{f}\:)\:h^{AC}\big)
h_{BD}\delta h^{BD}\ast 1\notag\\
&\:\:\:\:\:-\frac{1}{4}\int\nabla_{C}(\textit{f}\:K^{C})\ast
1-\frac{1}{4}\int\nabla_{B}\big(\nabla_{C}(\:\textit{f}\:)\delta
h^{CB}\big)\ast 1\notag\\
&\:\:\:\:\: -\frac{1}{2}\underset{\partial
N}{\int}dx^{(D-1)}\delta(\:\sqrt{|det\tilde{h}|}\:)N^{C}n_{C}.
\tag{\ref{app16}}
\end{align}
\end{subequations}
It is obvious that due to the vanishing of the variation of the
fields on the boundary $\partial N$ the last three terms above
which are the surface terms identically vanish. We remind the
reader of the action of the covariant derivative on the functional
$\textit{f}$
\begin{equation}\label{app17}
\begin{aligned}
\nabla_{B}(\:\textit{f}\:)&=\frac{\partial\:\textit{f}}{\partial\phi^{i}}\frac{\partial\phi^{i}}{\partial
x^{B}}
+\frac{\partial\:\textit{f}}{\partial\chi^{\alpha}}\frac{\partial\chi^{\alpha}}{\partial x^{B}},\\
\nabla_{A}\nabla_{B}(\:\textit{f}\:)&=\partial_{A}\partial_{B}\:\textit{f}-\partial_{C}\:\textit{f}\:\Gamma^{C}_{\:\:\:BA}.
\end{aligned}
\end{equation}
Thus in \eqref{app16} we may identify
\begin{equation}\label{app18}
\nabla_{C}\nabla^{C}(\:\textit{f}\:)\equiv\nabla_{C}\big(\nabla_{B}(\:\textit{f}\:)h^{BC}\big)=\big(\partial_{A}\partial_{B}\:\textit{f}-\partial_{C}
\:\textit{f}\:\Gamma^{C}_{\:\:\:BA}\big)h^{BA}.
\end{equation}


\begin{thebibliography}{99}
\bibitem{1}
Akama K., 1983 \textit{Pregeometry in Proceedings of the
International Symposium on Gauge Theory and Gravitation}, August
20–24, 1982, Nara, Japan, pp. 267.
\bibitem{2}
Rubakov V. A., and Shaposhnikov M. E. 1983 \textit{Phys. Lett.} B
\textbf{125} 136.
\bibitem{31}
Arkani-Hamed N., Dimopoulos S., Dvali G. 1998 \textit{Phys. Lett.}
B \textbf{429} 263 hep-ph/ 9803315.
\bibitem{32}
Arkani-Hamed N., Dimopoulos S., Dvali G. 1999 \textit{Phys. Rev.}
D \textbf{59} 086004 hep-th/9807344.
\bibitem{33}
Antoniadis I., Arkani-Hamed N., Dimopoulos S., and Dvali G. 1998
\textit{Phys. Lett.} B \textbf{436} 257 hep-ph/9804398.
\bibitem{34}
Randall L., Sundrum R. 1999 \textit{Phys. Rev. Lett.} \textbf{83}
3370 hep-ph/9905221.
\bibitem{35}
Randall L., Sundrum R. 1999 \textit{Phys. Rev. Lett.} \textbf{83}
4690 hep-th/9906064.
\bibitem{36}
Dvali G., Gabadadze G., and Porrati M. 2000 \textit{Phys. Lett.} B
\textbf{485} 208  hepth/ 0005016.
\bibitem{37}
Dvali G., and Gabadadze G. 2001 \textit{Phys. Rev.} D \textbf{63}
065007 hepth/0008054.
\bibitem{38}
Dvali G., Gabadadze G., Kolanovic M., and Nitti F. \textit{Phys.
Rev.} D \textbf{65} 024031 hep-th/0106058.
\bibitem{41}
Goldberger W. D., and Wise M. B. 1999 \textit{Phys. Rev. Lett.}
\textbf{83} 4922 hep-ph/9907447.
\bibitem{42}
Gremm M. 2000 \textit{Phys. Lett.} B \textbf{478} 434
hep-th/9912060.
\bibitem{43}
Gremm M. 2000 \textit{Phys. Rev.} D \textbf{62} 044017
hep-th/0002040.
\bibitem{44}
Csaki C., Erlich J., Hollowood T. J., and Shirman Y. 2000
\textit{Nucl. Phys.} B \textbf{581} 309 hep-th/0001033.
\bibitem{45}
Csaki C., Erlich J., Grojean C., and Hollowood T. J. 2000
\textit{Nucl. Phys.} B \textbf{584} 359 hep-th/0004133.
\bibitem{46}
Kobayashi S., Koyama K., and Soda J. 2002 \textit{Phys. Rev.} D
\textbf{65} 064014 hep-th/0107025.
\bibitem{47}
Parikh M. K., and Solodukhin S. N. 2001 \textit{Phys. Lett.} B
\textbf{503} 384 hep-th/0012231.
\bibitem{48}
Wang A. 2002 \textit{Phys. Rev.} D \textbf{66} 024024
hep-th/0201051.
\bibitem{49}
Minamitsuji M., Naylor W., and Sasaki M. 2006 \textit{Nucl. Phys.}
B \textbf{737} 121 hep-th/0508093.
\bibitem{410}
Bazeia D., Brito F. A., and Nascimento J. R. S. 2003 \textit{Phys.
Rev.} D\textbf{68} 085007 hep-th/0306284.
\bibitem{411}
Gogberashvili M., Singleton D. 2004 \textit{Phys. Lett.} B
\textbf{582} 95 hep-th/0310048.
\bibitem{412}
Bazeia D., and Losano L. 2006 \textit{Phys. Rev.} D \textbf{73}
025016 hep-th/0511193.
\bibitem{413}
Ghassemi S., Khakshournia S., and Mansouri R. 2007 \textit{Int. J.
Mod. Phys.} D \textbf{16} 629 gr-qc/0609132.
\bibitem{414}
Cvetic M., and Robnik M. 2008 \textit{Phys. Rev.} D \textbf{77}
124003 arXiv:0801.0801 [hep-th].
\bibitem{415}
Kanti P., Kogan I. I., Olive K. A., and Pospelov M. 1999
\textit{Phys. Lett.} B \textbf{468} 31 hep-ph/9909481.
\bibitem{416}
Mounaix p., and Langlois D. 2002 \textit{Phys. Rev.} D\textbf{65}
103523 gr-qc/0202089.
\bibitem{417}
Navarro I., and Santiago J. 2006 \textit{JCAP} \textbf{0603} 015
hep-th/0505156.
\bibitem{temel}
DeWolfe O., Freedman D. Z., Gubser S. S., and Karch A. 2000
\textit{Phys. Rev.} D \textbf{62} 046008 hep-th/9909134.
\bibitem{thick}
Dzhunushaliev V., Folomeev V., and Minamitsuji M. 2010
\textit{Rept. Prog. Phys.} \textbf{73} 066901 arXiv:0904.1775
[gr-qc].
\bibitem{sig1}
Coleman S. R., Wess J., and Zumino B. 1969 \textit{Phys. Rev.}
\textbf{177} 2239.
\bibitem{sig2}
Callan C. G., Coleman S. R., Wess J., and Zumino B. 1969
\textit{Phys. Rev.} \textbf{177} 2247.
\bibitem{sig3}
Gaillard M. K., and Zumino B. 1981 \textit{Nucl. Phys.} B
\textbf{193} 221.
\bibitem{westbook}
West P. C. 1990 \textit{Introduction to Supersymmetry, and
Supergravity} (2nd Edn.) (Singapore: World Scientific).
\bibitem{sssugradivdim}
Salam A., Sezgin E. 1989 \textit{Supergravities in Diverse
Dimensions Vol. 1, 2} (Singapore: World Scientific).
\bibitem{westsugra}
West P. C. 1998 hep-th/9811101.
\bibitem{tani}
Tanii Y. 1998 hep-th/9802138.
\bibitem{hel}
Helgason S. 2001 \textit{Differential Geometry, Lie Groups, and
Symmetric Spaces} (Providence R. I.: American Mathematical
Society).
\bibitem{sm1}
Eichenherr H., and Forger M. 1979 \textit{Nucl. Phys.} B
\textbf{155} 381.
\bibitem{sm2}
Eichenherr H., and Forger M. 1980 \textit{Nucl. Phys.} B
\textbf{164} 528.
\bibitem{sm3}
Eichenherr H., and Forger M. 1981 \textit{Commun. Math. Phys.}
\textbf{82} 227.
\bibitem{julia1}
Cremmer E., Julia B., L\"{u} H., and Pope C. N. 1998 \textit{Nucl.
Phys.} B \textbf{523} 73 hep-th/9710119.
\bibitem{julia2}
Cremmer E., Julia B., L\"{u} H., and Pope C. N. 1998 \textit{Nucl.
Phys.} B \textbf{535} 242 hep-th/9806106.
\bibitem{nej1}
Y$\i$lmaz N. T. 2003 \textit{Nucl. Phys.} B \textbf{664} 357
hep-th/0301236.
\bibitem{nej2}
Y$\i$lmaz N. T. 2003 \textit{Nucl. Phys.} B \textbf{675} 122
hep-th/0407006.
\bibitem{nej3}
Dereli T., and Y$\i$lmaz N. T. 2005 \textit{Nucl. Phys.} B
\textbf{705} 60 hep-th/0507007.
\bibitem{symmspace}
Y$\i$lmaz N. T. 2007 \textit{Int. J. Mod. Phys.} A \textbf{22}
2683 arXiv:0707.2150 [hep-th].
\bibitem{sssm1}
Y$\i$lmaz N. T. 2006 \textit{Phys. Lett.} B \textbf{642} 270
hep-th/0611010.
\bibitem{r1}
Gallot S., Hulin D., Lafontaine J 2004 \textit{Riemannian
Geometry} (Berlin: Springer-Verlag).
\bibitem{r2}
Petersen P. 1998 \textit{Riemannian Geometry} (Berlin:
Springer-Verlag).
\bibitem{r3}
Jost J. 2002 \textit{Riemannian Geometry, and Geometric Analysis}
(Berlin: Springer-Verlag).
\bibitem{thring}
Thirring W. 1992 \textit{A Course in Mathematical Physics I, and
II: Classical Dynamical Systems, and Classical Field Theory} (New
York: Springer-Verlag).
\bibitem{tucker}
Benn I. M., Tucker R. W. 1987 \textit{An Introduction to Spinors,
and Geometry with Applications in Physics} (Bristol: Adam Hilger).
\bibitem{nakahara}
Nakahara M. 1991 \textit{Geometry, Topology, and Physics}
(Bristol: Adam Hilger).
\bibitem{tem1}
Afonso V. I., Bazeia D., Menezes R., and Petrov A. Y. 2007
\textit{Phys. Lett.} B \textbf{658} 71
  arXiv:0710.3790 [hep-th].
  \bibitem{tem2}
Bazeia D., Brito F. A., and Losano L. 2006
  \textit{JHEP} \textbf{0611} 064 hep-th/0610233.
  \bibitem{branemot}
  Y$\i$lmaz N. T. 2010
  \textit{Class. Quant. Grav.} \textbf{27} 145019
  arXiv:1006.1968 [gr-qc].
\bibitem{baz10}
Bazeia D., Gomes A. R., Losano L., and Menezes R. 2009
  \textit{Phys. Lett.} B \textbf{671} 402
  arXiv:0808.1815 [hep-th].
\bibitem{baz1}
Bazeia D., Brito F. A., and Gomes A. R. 2004 \textit{JHEP}
\textbf{0411} 070 hep-th/0411088.
\bibitem{baz2}
Bazeia D., and Gomes A. R. 2004
  \textit{JHEP} \textbf{0405} 012 hep-th/0403141.
  \bibitem{baz3}
Bazeia D., Furtado C., and Gomes A. R. 2004
  \textit{JCAP} \textbf{0402} 002 hep-th/0308034.
  \bibitem{baz4}
Faulkner T., Horowitz G. T., and Roberts M. M. 2010
  \textit{Class.  Quant.  Grav.} \textbf{27} 205007
  arXiv:1006.2387 [hep-th].
  \bibitem{baz5}
Guerrero R., Melfo A., Pantoja N., and Rodriguez R. O. 2010
  \textit{Phys.  Rev.} D \textbf{81} 086004
  arXiv:0912.0463 [hep-th].
  \bibitem{baz6}
Zhao Z. H., Liu Y. X., and Li H. T. 2010
  \textit{Class. Quant. Grav.} \textbf{27} 185001
  arXiv:0911.2572 [hep-th].
  \bibitem{baz7}
Herrera-Aguilar A., Malagon-Morejon D., Mora-Luna R. R., and
Nucamendi U. 2010
  \textit{Mod. Phys. Lett.} A \textbf{25} 2089
  arXiv:0910.0363 [hep-th].
  \bibitem{baz8}
  Liu Y. X., Zhong Y., and Yang K. 2010
  \textit{Europhys. Lett.} \textbf{90} 51001
   arXiv:0907.1952 [hep-th].
\bibitem{fp}
  Fierz M. and Pauli W. 1939
   \textit{Proc. Roy. Soc. Lond.} A {\textbf173} 211.
\bibitem{ah}
  Arkani-Hamed N., Georgi H., and Schwartz M. D. 2003
   \textit{Annals Phys.}  {\textbf305} 96
  hep-th/0210184.
\bibitem{dgrt1}
  de Rham C., and Gabadadze G. 2010
   \textit{Phys. Rev.} D {\textbf 82} 044020
  arXiv:1007.0443 [hep-th].
\bibitem{dgrt2}
  de Rham C., Gabadadze G., and Tolley A. J. 2011
   \textit{Phys. Rev. Lett.}  {\textbf 106} 231101
  arXiv:1011.1232 [hep-th].
\bibitem{crem}
  Creminelli P., Nicolis A., Papucci M., and Trincherini E. 2005
   \textit{JHEP} {\textbf 0509} 003
  hep-th/0505147.
\bibitem{hr1}
  Hassan S. F., and Rosen R. A. 2011
   \textit{JHEP} {\textbf 1107} 009
  arXiv:1103.6055 [hep-th].
\bibitem{hr2}
  Hassan S. F., and Rosen R. A. 2012
   \textit{Phys. Rev. Lett.}  {\textbf 108} 041101
  arXiv:1106.3344 [hep-th].
\bibitem{aco}
  D'Amico G., de Rham C., Dubovsky S., Gabadadze G., Pirtskhalava D., and
  TolleyA. J. 2011
   \textit{Phys. Rev.} D {\textbf 84} 124046
  arXiv:1108.5231 [hep-th].
 \bibitem{tolley}
  Fasiello M., and Tolley A. J. 2012
   \textit{JCAP} {\textbf 1211} 035
  arXiv:1206.3852 [hep-th].
  \bibitem{ln}
  Langlois D., and Naruko A. 2012
   \textit{Class. Quant. Grav.}  {\textbf 29} 202001
  arXiv:1206.6810 [hep-th].
  \bibitem{has1}
  von Strauss M., Schmidt-May A., Enander J., Mortsell E., and Hassan S. F.
  2012
   \textit{JCAP} {\textbf 1203} 042
  arXiv:1111.1655 [gr-qc].
  \bibitem{has2}
  Hassan S. F., and Rosen R. A. 2012
   \textit{JHEP} {\textbf 1202} 126
  arXiv:1109.3515 [hep-th].
 \bibitem{comel}
  Comelli D., Crisostomi M., Nesti F., and Pilo L. 2012
   \textit{JHEP} {\textbf 1203} 067;
   Erratum-ibid.\ 2012 {\textbf 1206} 020
  arXiv:1111.1983 [hep-th].
  \bibitem{crisos}
  Babichev E., and Crisostomi M. 2013
  arXiv:1307.3640 [gr-qc].
\bibitem{khosra}
  Khosravi N., Rahmanpour N., Sepangi H. R., and Shahidi S. 2012
   \textit{Phys. Rev.} D {\textbf 85} 024049
  arXiv:1111.5346 [hep-th].
\bibitem{nomura}
  Nomura K., and Soda J. 2012
   \textit{Phys. Rev.} D {\textbf 86} 084052
  arXiv:1207.3637 [hep-th].
\end{thebibliography}
\end{document}